*PLM* modulates phloem unloading through sphingolipid biosynthesis and plasmodesmal ultrastructure

Dawei Yan*, Shri Ram Yadav*, Andrea Paterlini, William J. Nicolas, Ilya Belevich, Magali S. Grison, Anne Vaten, Leila Karami, Sedeer el-Showk, Jung-Youn Lee, Gosia M. Murawska, Jenny Mortimer, Michael Knoblauch, Eija Jokitalo, Jonathan E. Markham, Emmanuelle M. Bayer*, Ykä Helariutta*

## Abstract

During phloem unloading, multiple cell-to-cell transport events move organic substances to the root meristem. Whereas the primary unloading event from the sieve elements (SE) to the phloem pole pericycle (PPP) has been characterized to some extent, little is known about post-SE unloading. Here, we report a novel gene, *PLM (PHLOEM UNLOADING MODULATOR)*, in the absence of which plasmodesmata-mediated symplastic transport through the PPP-endodermis interface is specifically enhanced. Increased unloading is attributable to a defect in the formation of the ER-plasma membrane tethers during plasmodesmal morphogenesis, resulting in the majority of pores lacking a visible cytoplasmic sleeve. *PLM* encodes a putative enzyme required for the biosynthesis of sphingolipids with very long chain fatty acid (VLCFA) chains. Taken together, our results indicate that post-SE unloading involves sphingolipid metabolism which impacts plasmodesmal ultrastructure. They also raise the question of how and why plasmodesmata with no cytoplasmic sleeve facilitate molecular trafficking.

## Main

Plasmodesmata are membrane-lined channels that cross the plant cell wall, connecting neighboring cells to mediate symplastic communication[1]. Cell-to-cell trafficking of a wide range of molecules via plasmodesmata is involved in the coordination of growth and developmental programs such as cell differentiation, photoassimilate translocation, disease and stress resistance[2-9]. The typical plasmodesmal structure consists of the plasma membrane (PM) lining the pore and a central rod-like structure, called the desmotubule, derived from the endoplasmic reticulum (ER )[10-12]. Inside the pore, the ER and the PM are tethered by unidentified spoke-like elements, and this specialized membrane arrangement defines plasmodesmata as a specialized type of membrane contact site. The gap between the PM and the desmotubular ER is termed the cytoplasmic sleeve, the presence of which classifies plasmodesmata as type I (no visible sleeve) or type II (clear cytoplasmic sleeve)[13]. In addition, plasmodesmata can also be classified as simple (with only a single channel) or branched (with multiple channels joining into a central cavity) based on their morphology[14,15]. In current models, the cytoplasmic sleeve is assumed to facilitate molecular trafficking through plasmodesmata[16,17]. The larger the gap between the ER and the PM, the more open the cytoplasmic sleeve and thus the larger the size exclusion limit of the pores. However, this hypothesis was recently challenged by a report that type I plasmodesmata may predominate at cellular interfaces with efficient symplastic trafficking and also allow the movement of both micromolecules and macromolecules[13].

Plasmodesmata consist of several chemically distinct components. Callose ($\beta$-1,3-glucan) accumulates in the cell wall around the pores, and its abundance is known to inversely correlate with trafficking efficiency[18-20]. Several membrane protein classes, including receptor-like proteins, are known to be associated with plasmodesmata[21-23]. The membrane lipid environment of the pores is distinct from that of the bulk PM, with an enrichment of complex sphingolipids with saturated very long chain fatty acids (VLCFAs) and a higher ratio of sterols to glycerolipids[24]. Sterols play a role in the regulation of plasmodesmatal permeability, and inhibiting sterol synthesis affects the targeting of plasmodesmata-localized proteins and ultimately callose homeostasis[24,25]. However, the precise role of sphingolipids in plasmodesmata remains elusive.

As discussed above, structurally diverse plasmodesmata have been identified and the abundance of the various plasmodesmal forms appears to vary depending on the cellular interface. Recently, another structural variant, funnel plasmodesmata, was identified in the context of phloem unloading[26]. Transmission electron microscopy revealed large V-shaped structures in SE cell walls, which appear to unload especially large proteins in batches to the neighboring PPP[26], indicating a structural adaptation of the symplastic pathway for phloem unloading.

Here, we identify *PLM (PHLOEM UNLOADING MODULATOR)* as a modulator of symplastic unloading at the interface between the PPP and endodermis in roots. We show that PLM is involved in sphingolipid metabolism and affects plasmodesmal structure. In Arabidopsis, *plm* loss-of-function mutants present a defect in the ER-PM tethering transition from type I to type II plasmodesmata, which is correlated with increase trafficking at the PPP-endodermis interface. Our data identify sphingolipids as regulators of plasmodesmal structure and unexpectedly reveal that plasmodesmata with a tight cytoplasmic sleeve are more competent at trafficking than those without a sleeve.

## Results

**Isolation and identification of the *plm* mutation, a suppressor of gain-of-function alleles in callose biosynthesis**

We have previously identified gain-of-function mutants (*cals3-d*) of the *CALLOSE SYNTHASE 3/GLUCAN SYNTHASE-LIKE12 (CALS3/GSL12)* gene[27]. The *cals3-d* mutants had reduced plasmodesmatal permeability, which in turn resulted in decreased intercellular trafficking, impaired phloem unloading and defective root development in Arabidopsis[27]. To further investigate the genetic control of symplastic trafficking, a genetic screen for suppressors was performed using EMS mutagenized *cals3-1d* plants expressing *pSUC2::GFP*[27]. In addition to numerous intragenic mutants, a putative extragenic suppressor of *cals3-1d* was eventually identified based on GFP unloading and root growth. The suppressor partially rescued the phloem unloading and root phenotypes of *cals3-1d*. GFP transport was restored up to the transition zone, and seedling growth was similar to wild-type C24 (Fig. 1a and Supplementary Fig.1b). A genetic linkage analysis of F2 individuals derived from a cross between the suppressor and *cals3-3d* suggested that GFP trafficking and root growth were influenced by a single recessive mutation which we named *plm-1*. By PCR-based positional cloning, the *PLM* locus was narrowed down to a window of approximately 204 kb in chromosome 1. Subsequent whole-genome resequencing identified a premature STOP codon (TGG to TGA, 717 position of CDS) in exon 3 of the gene *At1g43580* (Fig. 1b). When complemented with a 3,705 bp genomic DNA fragment which included 1,508 bp upstream of the gene, the suppressor line (*cals3-1d;plm-1*) displayed GFP unloading defects and a short primary root similar to *cals3-1d* (Fig.1a and Supplementary Fig.1b), supporting the identification of *At1g43580* as *PLM*.

To further study the function of *PLM*, we isolated the *plm-1* mutant in the C24 background. Using *pSUC2::GFP* as a reporter, we found a significant increase in the GFP signal in the root meristem of *plm-1* (Fig.1d). To confirm the relationship between the enhanced GFP-trafficking phenotype and the *plm* mutation, an independent knock-out T-DNA insertion mutant (SALK_064909) in Col-0 was identified, which we named *plm-2* (Fig. 1b,c). Like *plm-1*, *plm-2* displayed enhanced GFP-trafficking (Fig. 1e). Furthermore, the *plm-2;cals3-3d* double mutant with *pSUC2::GFP* showed a similar GFP trafficking pattern to *plm-1;cals3-1d* and could also be complemented by genomic *PLM* and a *PLM-GFP* fusion (Supplementary Fig. 1a,c). We next characterized the overall phenotype caused by the recessive *plm* mutations. Both *plm-1* and *plm-2* displayed an early spurt in root elongation, conferring a slight but significant difference in root elongation for several days (Fig. 1f,g and Supplementary Fig.2a). We also found that the enhanced GFP-unloading phenotype in *plm-2* was evident by day 3 after germination (Supplementary Fig. 2f), coinciding with the early root elongation phenotype (Supplementary Fig. 2a). Otherwise, we did not detect other phenotypic differences; for example, general growth and biomass accumulation were similar in the mutants and wild type (Supplementary Fig. 2b,c,d,e).Taken together, our results indicate that *PLM* influences the root elongation rate by controlling the extent of phloem unloading soon after germination.

**PLM encodes a novel protein required for accumulation of VLCFA-containing sphingolipids**

Based on the Arabidopsis database (http://www.Arabidopsis.org), PLM is a single-copy gene that contains 4 exons and 3 introns. It is predicted to encode a sphingomyelin synthase (SMS) family protein with a weak phosphatidic acid phosphatase-related2 domain (PAP2_C) (Fig.2a). Sphingomyelin is an important type of sphingolipid found in organisms ranging from protozoa to mammals[28,29]. It is synthesized by the transfer of phosphocholine from phosphatidylcholine (PC) to ceramide, yielding diacylglycerol (DAG) as a side product[30,31]. The reaction is catalyzed by SMS, which is located at the PM and in the Golgi in most mammalian cells[32,33,34]. However, sphingomyelin has not been detected in plants, as confirmed by our LC-MS analysis in Col-0 Arabidopsis (data not shown). However, plants possess inositol phosphorylceramide synthase (IPCS) enzymes, which also have a PAP2_C domain and employ ceramides as substrates to catalyse the synthesis of IPC in the Golgi[34].

Multiple sequence alignment of full-length protein sequences showed that PLM has very low sequence identity (20-26%) with both human SMSs and Arabidopsis IPCSs (Supplementary Fig. 3 and 4). PLM has the typical D1 domain of SMS proteins but not the D2 domain. It possesses the three conserved amino acid residues H-H-D in the catalytic site for phosphatidic acid phosphatase activity while displaying low identity for other residues in the D3 and D4 domains[32] (Supplementary Fig. 3). A phylogenetic tree of known and predicted SMSs, IPCSs and PLM homologues from various organisms was generated. It revealed that PLM is conserved among vascular plants and is distinct from SMSs and IPCSs (Fig.2b).

A *pPLM::PLM-YFP* construct was employed to investigate the expression pattern and subcellular localization of PLM. The construct could revert the unloading phenotype of *plm-2;cals3-3d* back to *cals3-3d* levels, demonstrating that it was functional (Supplementary Fig.1a). *pPLM::PLM-YFP* displayed a broad expression pattern in the root (Fig.2c). The tagged protein co-localized with an ER-retained RFP reporter, indicating that PLM is ER-localized (Fig.2d).

To investigate the biochemical basis for the *plm* phenotype and the role of *PLM* in sphingolipid metabolism, a complete sphingolipid analysis was performed on 5-day old root seedlings[35]. The four major classes of sphingolipids (ceramides, hydroxyceramides, glucosylceramides, and glycosyl inositol phospho ceramides (GIPCs)) were analyzed along with free long chain bases (LCBs) and phosphorylated LCBs (LCBPs), which are sphingolipid precursors (Fig. 3a,b). A comparison between Col-0 and *plm-2* showed that the trihydroxy LCBs t18:0 and t18:1, the main precursors of GIPCs, were decreased up to 37.8% and 28.9% in the mutant, respectively, while dihydroxy LCB species (d18:0 and d18:1) were unaltered (Fig. 3c). Consistent with this, a significant reduction not only in ceramides but also in GIPC species containing VLCFA was observed in *plm-2*: 24:0-ceramide (t18:0, decrease of 48.3% and t18:1, decrease of 48.9%), 24:0-GIPC (t18:0, decrease of 20.7% and t18:1; decrease of 24.1%) (Fig. 3d). No detectable alteration in species containing shorter 16:0 FA was found (Fig. 3d). Some VLCFA-containing hydroxyceramides were also increased, while most VLCFA-containing glucosylceramides were unaltered (Fig. 3d). Overall, total ceramide and GIPC levels were significantly reduced in *plm-2*, by 34.6% and 22.0%, respectively (Fig. 3e).

GIPCs are the most abundant and complex sphingolipid species *in Arabidopsis*. They possess different head groups that are modified by a number of sugar residues[36,37]. To investigate whether *plm*-2 has a depletion of specific GIPC head species, primuline and orcinol staining in addition to MALDI analysis were performed. No differences were found (Supplementary Fig. 5), suggesting a similar composition of GIPC sugar heads in Col-0 and *plm-2*.

To investigate whether PLM is specifically involved in sphingolipid metabolism, profiling of polar lipid, DAG and sterol content was performed. No alterations in the amounts of these lipids were detected in *plm-2* (Supplementary Fig. 6a,b,c). Overall, these results demonstrate that the *plm* mutation interferes specifically in sphingolipid metabolism, mainly resulting in reduced tri-hydroxylated LCBs and VLCFA-GIPC sphingolipids.

To further strengthen the difference between PLM and IPCS2, a T-DNA insertion allele of *Atipcs2* (SALK_206784) was also included in the sphingolipid profiling. The relative subtle changes observed in *ipcs2* mutant (Supplementary Fig. 7) did not match the more significant ones in *plm-2.* Based on the phylogenetic clustering of

*plm* and its difference from IPCSs in terms of subcellular localization and sphingolipid profiles, PLM is likely to be a new plant-specific enzyme.

**plm enhances GFP post-SE unloading from PPP to endodermis**

We next focused on the unloading phenotype of *plm-2*. As described above, *plm* mutations enhance GFP trafficking from the phloem to the root tip. We therefore explored the unloading process in more detail. GFP expressed by *pSUC2::GFP* translocates from companion cells to SEs then unloads symplastically into the root meristem from protophloem SEs through the PPP-endodermis interface (Fig. 4a). While wild-type plants presented a strong difference in GFP signal between the unloading domain in the transition zone and the root meristematic zone, *plm-2* has a less-defined boundary and a more uniform GFP signal (Fig. 1d,e). GFP unloading in the root tip therefore appears to be enhanced in *plm-2*. We confirmed this using an inducible version of *pSUC2::GFP*, which displayed a similar difference between unloading in wild type and *plm-2* upon induction (Supplementary Fig. 8a). The unaltered transcript levels of *GFP* and *SUC2* excluded the possibility of regulation of *SUC2* expression by *PLM* (Supplementary Fig.8b). Furthermore, no developmental differences in vasculature were observed in *plm-2* based on established markers for phloem (*pAPL::YFPer*), SE (*pCALS7::H2B-YFP*) and PPP (*pCALS8::YFPer*) (Supplementary Fig.9).

Phloem unloading in the root tip occurs from proto-SEs into PPP cells. Further subsequent radial movement into neighboring endodermal cells is tightly regulated and depends on the size of the unloaded macromolecules[26] (Fig.4a). In wild-type plants, PPP cells display a stronger GFP signal intensity than endodermal cells (Fig.4b,c). The signal maximum suggests that GFP experiences some form of bottleneck in transport at the PPP-endodermis interface. Differences in the GFP signal between the two layers were reduced in *plm-2* (Fig.4b,c,d), indicating that the trafficking restriction that normally occurs at the PPP-endodermis interface is relaxed in the *plm-2* mutant. To further explore GFP trafficking across the PPP-endodermis interface, fluorescence recovery after photobleaching (FRAP)[38] was performed in the unloading zone. The GFP recovery rate in the endodermis with influx from PPP was faster in *plm-2* than in Col-0, while the movement through the

SE-PPP interface was not affected (Fig.4e,f). These data together suggest that post SE-PPP unloading trafficking is affected in *plm-2*, with a specific impact on the PPP-endodermis interface.

To strengthen the link between sphingolipid homeostasis by *PLM* and enhanced GFP trafficking, we used the ceramide synthase inhibitor FB1, which reduces ceramide and GIPC products[39,40]. After a 72h treatment at concentrations shown not to affect root growth, the GFP signal in the root tip of Col-0 plants carrying *pSUC2::GFP* was increased, mimicking the *plm* mutation (Supplementary Fig.10a,b).

**PLM-mediated symplastic trafficking is independent from plasmodesmata density and callose accumulation**

We tested multiple hypotheses to determine which change at the PPP-endodermis interface underpins the enhanced GFP trafficking in *plm-2*. The density of plasmodesmata in the unloading zone was first assessed employing serial block face scanning electron microscopy (SB-EM). There was no increase in the number of plasmodesmata per unit area at the SE-PPP or PPP-endodermis interfaces of *plm-2* (Fig. 4g), indicating that the enhanced trafficking is not the result of increased pores.

Since callose is known to modulate symplastic trafficking by regulating plasmodesmata aperture and *plm* is a suppressor of *cals3d*, we analyzed the transcript levels of *CALS3* in the root and the subcellular localization of the CALS3 protein. Both were unaltered in *plm-2* (Fig. 4h,i). To assess the callose level, immunolocalization using a callose antibody on root sections from the unloading domain was performed. No difference was observed in the callose signal at the PPP-endodermis interface of Col-0 and *plm-2* (Fig.4j). In summary, our results indicate that the *plm* mutation enhances plasmodesmatal permeability independently of plasmodesmata density and callose regulation.

To further investigate the impact of callose on *PLM*-mediated trafficking, we carried out an assay of root growth under stress caused by alloxan treatment, which induces callose accumulation and blocks phloem unloading, compromising root growth[41]. At 28h, 52h and 65h after transferring 5-day old seedlings to alloxan media, *plm-2* roots grew longer than wild-type roots, while no difference was observed after transferal to normal media (Supplementary Fig. 11). This result suggests that *plm* plasmodesmata are somewhat resistant to the negative effect of callose deposition. It is possible that the ability of the *plm* mutation to suppress callose accumulation in *casl3-d* is based on this characteristic.

## *PLM* regulates plasmodesmal architecture at the PPP-endodermis interface

One possible explanation for the observed enhanced symplastic conductivity is an alteration in plasmodesmal ultrastructure. Nicolas et al (2017)[13] have shown that type I plasmodesmata (lacking an electron-lucent cytoplasmic sleeve due to close PM-ER contact) are predominant in young root tissue and open to macromolecular trafficking, while type II plasmodesmata (displaying a visible electron-lucent cytoplasmic sleeve with a larger ER-PM gap containing sparse tether-like spoke elements) are more numerous in older tissue. To investigate possible alterations in plasmodesmal ultrastructure, we employed electron tomography[13] at the PPP-endodermis interface in Col-0 and *plm-2*. Detailed examination of our samples showed that while Col-0 plants had an equal proportion of type I and type II plasmodesmata at the PPP-endodermis interface, *plm-2* had no type II plasmodesmata (Fig. 5a,b). We also detected a few intermediates, with no clear spokes but partial detachment between the two membranes (Fig. 5b). In terms of morphology, a mixed population of simple and branched plasmodesmata were observed in both Col-0 and *plm-2*. In Col-0, the branched plasmodesmata were all type II (presumably corresponding to later maturation events), while simple plasmodesmata consisted of both type I and type II (Fig. 5a,b). Defective transition from type I to type II plasmodesmata in the *plm-2* mutant was also demonstrated by measuring the width of the pores (type I being narrower than type II). In Col-0, plasmodesmatal width increased between simple type I and branched type II plasmodesmata (Fig. 5c). By contrast, there was no clear difference in the width of simple and branched plasmodesmata in the *plm-2* mutant (Fig. 5c). The loss of the width increase between simple and branched plasmodesmata mirrored the depletion of type II plasmodesmata. Overall, the *plm* mutant therefore seems to be compromised in the transition from type I to type II plasmodesmata but not from simple to branched structures.

## Discussion

In this paper, we establish a link between sphingolipid metabolism, the internal architecture of plasmodesmata and cell-to-cell connectivity. Our results provide evidence for an unexpected role of the cytoplasmic sleeve and ER-PM contacts in plasmodesmata-mediated cell-to-cell trafficking.

The cytoplasmic sleeve of plasmodesmata has been well established as a continuum of symplasm between adjacent cells, and the size exclusion limit of the pores is believed to depend on the conductivity of the cytoplasmic sleeve[16,17]. Recently Nicolas et al. (2017) postulated a positive correlation between the spatial distribution of type I plasmodesmata (with a very narrow cytoplasmic sleeve) and enhanced trafficking[13]. Here, we provide genetic support for this model and furthermore demonstrate that type I plasmodesmata are more conductive than type II, which have an "open" cytoplasmic sleeve. The lack of type II plasmodesmata and the higher GFP conductivity at the PPP-endodermis interface in *plm* (together with the unchanged density of plasmodesmata and callose level) strongly indicate that the narrow cytoplasmic sleeve observed in type I plasmodesmata may facilitate symplastic trafficking. However, how and why type I plasmodesmata favor cell-to-cell transport remains elusive.

Because of the structural homology of PLM with the SMSs and IPCSs, which are involved in sphingolipid metabolism, we profiled sphingolipids and other lipids in the *plm-2* mutant. We observed a specific effect of the *plm* mutation on sphingolipids. Phospholipids, DAG or sterols were not affected. Most striking was the substantial reduction in trihydroxy LCBs (but not dihydroxy LCBs), as well as the downstream ceramide and GIPC products, especially those containing VLCFA (Fig. 3). The profile seems unique compared to known sphingolipid biosynthetic mutants. Very few mutants in the sphingolipid biosynthesis pathway show a reduced level of total sphingolipids without also showing a severe growth phenotype, further indicating the uniqueness of the *plm* mutation. Based on sequence alignment and sphingolipid profiling, it is possible that *PLM* functions as an enzyme at an unknown synthesis branch for trihydroxy LCB synthesis in the ER. Alternatively, PLM may indirectly affect the sphingolipid biosynthesis pathway and act as a regulatory element.

The question of how sphingolipids affect plasmodesmata remains unresolved. Sphingolipid-containing VLCFAs and especially GIPCs have been reported to be critical for membrane bending and microdomain maintenance and formation[42-45]. In plants, VLCFA-sphingolipids have been involved in tissue polarity and protein sorting[46,47]. Finally, GIPCs have been found to be enriched in plasmodesmata-enriched fractions[24]. One scenario is that sphingolipids are required for the establishment of plasmodesmata-specific microdomains and affect the localization of structural components, the physical properties of the pores or the immediate cell wall environment, all of which could be required for the transition from type I to type II.

In this work, we provide further elements to dissect the phloem unloading pathway. Our previous collaborative work revealed that the SE-PPP route is the predominant unloading pathway[26]. Small molecules such as sugar and GFP can be unloaded from SE into PPPs and then enter the post-SE unloading pathway to the meristem, while larger proteins (for instance, the size of SE OCCLUSION RELATED (SEOR) protein fused to YFP (SEOR-YFP)) remain trapped within the PPP[26]. Therefore, the post-phloem PPP-endodermis interface has emerged as a key barrier for controlling post-unloading transport to the sinks. The higher conductivity of the PPP-endodermis interface and enhanced GFP input into the root meristem are consistent with the accelerated root elongation seen in young *plm* seedlings, indicating a transient, rate-limiting role for the PPP-endodermis interface in symplastic trafficking. However, the enhancement of plasmodesmatal permeability by PLM is most likely limited, as larger proteins, such as SPORAMIN-GFP and SEOR-YFP, are retained in the PPP in *plm* (Supplementary Fig. 12). This suggests the existence of other regulatory mechanisms which might contribute to the somewhat subtle nature of the phenotype in Arabidopsis.

A unique type of funnel-shaped plasmodesmata has been reported at the SE-PPP interface, indicating symplastic interfaces are specialized. The observed change in conductivity, based on FRAP, at the PPP-endodermis interface but not at the SE-PPP interface suggests further specialization in the unloading interfaces. Indeed, although PLM is expressed broadly during root development, it is possible that the requirement for VLCFA-containing sphingolipids is quite local. Tissue-specific relevance of VLCFA-containing lipids in the pericycle has been reported in the context of callus generation[48].

Finally, we found *PLM*-like genes in the genomes of virtually all vascular plants that we investigated (Fig.2b), but how the *PLM* pathway contributes to the functional diversity of symplastic trafficking events in plants remains to be assessed.

## Methods

### Plant materials and growth conditions

*Arabidopsis thaliana* mutants *cals3-1d* and *plm-1* are in the C24 background; *cals3-3d* and *plm-2* are in the Col-0 background. *plm-1* was isolated in a genetic screen based on EMS mutagenesis of a gain-of-function allele of *cals3-1d* to identify a second site mutation that could fully or partially rescue root growth and *pSUC2::GFP* unloading defects. The segregation pattern of identified putative suppressors was analyzed in the M3 generation to determine whether the suppressor was dominant or recessive. Next, suppressors were backcrossed with the *pSUC2::GFP* line (C24 background), and the phenotypic segregation pattern was studied in the F1 and F2 generations to identify putative extra-genic suppressors. A mapping population was generated by crossing a non-segregating *plm* suppressor mutant (C24) with *cals3-3d* (Col-0). A PCR-based positional cloning was used for rough mapping, and whole-genome re-sequencing (Illumina) was used for SNP identification. *plm-2* (SALK_064909) was obtained from the Nottingham Arabidopsis Stock Centre. Seedlings were sterilized and grown vertically in Petri dishes on 1/2 Murashige and Skoog (MS) basal salt mixture with 1% agar, 1% sucrose and 0.05% 4-morpholine ethanesulfonic acid (MES) at pH 5.8 in a growth chamber at 23°C under 16h light conditions.

**Transgenic work**

Transgenic constructs were generated using the Gateway or MultiSite Gateway system (Invitrogen) and were introduced into plants using the floral dip method[49]. For *pPLM::PLM-YFP*, a 2.2-kb genomic DNA fragment with a 1.5-kb upstream region was amplified and combined with YFP into the MultiSite destination vector pBm43GW. Analysis was carried out in the F2 or F3 generations of crosses with wild type or mutant plants. Primer sequences are listed in Supplemental Table 1.

**Confocal microscopy**

Confocal imaging was performed on a Leica TCS SP8. Excitation and emission spectra were, respectively, 484 and 489-505 nm for GFP, 514 and 522-564 nm for YFP, 561 and 580-620 for RFP, and 561 and 600-650 for propidium iodide (PI). PI was used to stain cell contours.

**Transcript analysis**

For quantification of mRNA, root tips from 5-day-old seedlings were harvested. The total RNA was extracted with a miRNeasy Mini Kit (Quiagen) and treated with DNase on a column (QIAGEN). cDNAs were synthesized from total RNA primed with oligo(dT)18 primers using a First strand synthesis kit (Roche). Semi-quantitative PCR for *plm-2* was performed using specific primers across the T-DNA insertion site and *ACTIN2* as an internal standard. Quantitative PCRs (qPCRs) were performed using gene-specific primers and real-time PCR mix (Roche) in a LightCycler 480 Real-Time PCR System (Roche) with a standard program for 40 cycles. The levels of expression were calculated relative to *ACTIN2*. Primer sequences are provided in Supplemental Table 1.

**Sphingolipid profile**

Sphingolipids were extracted from 10 to 30 mg of lyophilized 5-day-old roots. The molecular species of different sphingolipid classes were analyzed by reverse-phase HPLC coupled to ESI-MS/MS as described in[35].

**FRAP assay**

FRAP was conducted on a Leica TCS SP8 confocal microscope. GFP in the layers outwards of the SE or PPP in the unloading zone was photo-bleached with 476,

488 and 496 nm lasers concurrently with 100% power and 8x zoom with a 25X lens. Subsequent recording of the refilling occurred immediately following photobleaching at 0.75X digital zoom by excitation with the 488 nm laser line, and emission was collected between 505 and 545 nm. The relative recovery profiles were normalized to pre-bleach image with the first image post-bleach set to 0.

**SB-EM**

Five-day old Col-0 and *plm-2* roots were fixed, dehydrated and embedded in Durcupan resin following protocols previously described in[50,51] . To improve handling, the roots were shortened to a total length of about 600 µm from the tip using a grinding machine (LaboPol-5, Struers Inc). Ground roots were then mounted by the ground side to a 3View pin and trimmed from the opposite end to 240 µm from the root tip using an EM Ultracut UC6i ultramicrotome (Leica Mikrosysteme GmbH) and a diamond knife. The trimming distance was selected based on markers for SE enucleation to get as close as possible to the area of interest. As a control, a histological section was cut and checked after the trimming. The blocks were covered with silver paint (Agar Scientific Ltd.) and platinum-coated using Quorum Q150TS (Quorum Technologies, UK). Images were acquired with an FEG-SEM Quanta 250 (FEI, Hillsboro, OR) using a backscattered electron detector (Gatan Inc.) with 2.5- kV beam voltage, spot size 3 and pressure between 0.1 to 0.3 Torr. Images were collected moving away from the stumped root tip in 40 nm steps. The coordinates of bonding boxes for each dataset were recalculated from root tip for ease of interpretation.

The datasets were processed using the Microscopy Image Browser (MIB)[52], run within the Matlab environment (Mathworks Inc.). Plasmodesmata were manually annotated on the cell-cell interfaces of interest. Care was taken to not count the same PD twice, making a conservative estimate. Midlines were drawn across the walls corresponding to the interfaces of interest on multiple slides. Lines were then interpolated and a 3D surface generated using the specially developed Contact Area 3D plugin of MIB. Density was subsequently calculated as the number of plasmodesmata per unit of surface.

## Callose immunofluorescence localization

Callose detection was conducted according to[27]. Roots of 5-day-old seedlings were fixed using 4% formaldehyde (Sigma) and 0.5% glutaraldehyde (Sigma) in 1x PBS buffer at pH7 overnight at 4°C. Roots were then washed, aligned and embedded in 1% low-melting Agarose (Calbiochem). After dehydration by 25%, 50%, 70%, 90%, 96%, 3 × 100% ethanol and resin infiltration (LR White medium grade, Agar scientific), polymerization was conducted overnight at 60°C. Callose immunolocalization was performed on 1 µm sections collected within a 100 µm unloading domain starting from the first enucleated SE using monoclonal antibody to (1→3)-β-glucan (Biosupplies, AU) as the primary antibody (1:1000) and Alexa Fluor Plus 647 anti-mouse IgG as a secondary antibody (1:1000). Slides were finally mounted in a 1:1 solution of AF1 antifading agent (Citifluor) and 1x PBS, with the addition of calcofluor as a cell wall counterstain. Images were taken by confocal laser scanning microscopy (Leica TCS SP8) and processed in Image J (https://imagej.nih.gov/ij/).

## Electron tomography

Sample preparation, sectioning, TEM analysis, tomogram reconstruction and analysis were performed as described[13,53].

High pressure freezing-freeze substitution: Root tips from Col-0 and *plm-2* mutant plants were frozen with high pressure to achieve vitreous ice in the whole sample. 5-day old seedlings were placed on a glass microscopy slide covered with 20% BSA. Root tips were sampled over 1.5 mm approximately from the root tip in order to have access to the differentiation zone, where phloem enucleation and unloading occur. Roots were then placed in Leica copper membrane carriers with a drop of BSA 20% to serve as cryoprotectant and filler. The loaded membrane carriers were then processed in an EM PACT1 (Leica) high pressure freezer, with no more than 30 seconds separating sampling and freezing. Samples were stored at -196°C in liquid nitrogen until further processing. Samples were then transferred to the well of the AFS2 machine (Leica), precooled to -90°C for the freeze substitution steps. The frozen membrane carriers were incubated in sealed cryotubes with a cryosubstitution cocktail containing 2% osmium tetroxide, 0.1% uranyl acetate and 0.5% glutaraldehyde in pure acetone. The cocktail was left at -90°C for 48h and then temperature was progressively brought up to -50°C (+3°/h). The cryofixation cocktail was then carefully removed and washed out by three consecutive 10-minute long acetone baths. Three pure ethanol baths were then performed, and uncasing of the samples from the copper membrane carriers was performed at the last one. Samples were carefully placed in resin casts (either coffin shaped ones for transversal root sections or cylindrical ones for longitudinal sections). Incubation of the samples in solutions of increasing concentration of HM20 Lowicryl resin in pure ethanol was as follows: 2h in 25% and 50% HM20, overnight in 75% HM20. This was followed by two consecutive 2h baths in 100% HM20. Curing the resin was accomplished by a last incubation in 100% HM20 for 8h before exposing the samples to UV light for 48h using the FSP module of the AFS2 (Leica). The first 24h of UV exposure were done at -50°C, then the temperature was raised to room temperature for the last 24h of UV exposition.

Section collection: Blocks were cut using a EM UC7 ultramicrotome (Leica). Section thickness ranged from 90 to 150nm. Sections were deposited on either formvar filmed and carbon coated slot grid (Electron Microscopy Science) or on 200-mesh copper grids coated with 2% parlodion. Grids with deposited sections were incubated in a solution of 5nm gold fiducial markers on both sides of the grid. The solution is a colloidal gold solution (BBI solutions) mixed with 0.5% BSA in a 1:1 ratio.

Data acquisition: Tilt series were acquired on a FEI TECNAI Spirit 120kV electron microscope equipped with a -70° - +70°C tilting goniometer and an Eagle 4k x 4k bottom camera. A tomography optimized single tilt holder was used (Fichione instruments, model-2020). Acquisitions were done at magnifications between 30 000x and 56 000x with a 1° increment for a range of -65° to +65°. Dual tilt axis tomography was achieved by taking out the sample holder and manually rotating the grid approximately 90° before acquiring the other axis. The batch mode of FEI 3D-explore was used to acquire tilt series automatically on areas of interest.

Tomogram reconstruction: The raw unbinned 4k x 4k data was processed using the IMOD package with the etomo GUI[54]. Tilt series from both axis were binned down to 2k x 2k to reduce occupancy and processing time for alignment and reconstruction. Alignment was performed using the automatic bead detection/tracking script, and reconstruction was done by weighed back projection, using the SIRT-like filter set at an equivalent of 15 iterations, or by real SIRT, 15 iterations. Then the two tomograms from the orthogonal axis were combined using IMOD again.

Data segmentation: Segmentation was done manually using 3dmod from the IMOD package. The interpolator and diverse drawing tools were used to trace the multiple features of interest. Systematic measurements were taken on all relevant unfiltered tomograms.

**Reporting Summary**

Further information on the experimental design is available in the Nature Research Reporting Summary linked to this article.

**Data availability**

All data underlying the findings are available from the corresponding author upon request.

# References

1   Lucas, W. J. Plasmodesmata - Intercellular Channels for Macromolecular Transport in Plants. *Curr Opin Cell Biol* **7**, 673-680, doi:Doi 10.1016/0955-0674(95)80109-X (1995).
2   Kim, I. & Zambryski, P. C. Cell-to-cell communication via plasmodesmata during Arabidopsis embryogenesis. *Curr Opin Plant Biol* **8**, 593-599, doi:10.1016/j.pbi.2005.09.013 (2005).
3   Otero, S., Helariutta, Y. & Benitez-Alfonso, Y. Symplastic communication in organ formation and tissue patterning. *Curr Opin Plant Biol* **29**, 21-28, doi:10.1016/j.pbi.2015.10.007 (2016).
4   Lim, G. H. *et al.* Plasmodesmata Localizing Proteins Regulate Transport and Signaling during Systemic Acquired Immunity in Plants. *Cell Host Microbe* **19**, 541-549, doi:10.1016/j.chom.2016.03.006 (2016).
5   Yadav, S. R., Yan, D., Sevilem, I. & Helariutta, Y. Plasmodesmata-mediated intercellular signaling during plant growth and development. *Front Plant Sci* **5**, 44, doi:10.3389/fpls.2014.00044 (2014).
6   Benitez-Alfonso, Y. *et al.* Symplastic intercellular connectivity regulates lateral root patterning. *Dev Cell* **26**, 136-147, doi:10.1016/j.devcel.2013.06.010 (2013).

7       Sivaguru, M. *et al.* Aluminum-induced 1-->3-beta-D-glucan inhibits cell-to-cell trafficking of molecules through plasmodesmata. A new mechanism of aluminum toxicity in plants. *Plant Physiol* **124**, 991-1006 (2000).
8       Tylewicz, S. *et al.* Photoperiodic control of seasonal growth is mediated by ABA acting on cell-cell communication. *Science* **360**, 212-214, doi:10.1126/science.aan8576 (2018).
9       Oparka, K. J. *et al.* Simple, but not branched, plasmodesmata allow the nonspecific trafficking of proteins in developing tobacco leaves. *Cell* **97**, 743-754 (1999).
10      Tilsner, J., Amari, K. & Torrance, L. Plasmodesmata viewed as specialised membrane adhesion sites. *Protoplasma* **248**, 39-60, doi:10.1007/s00709-010-0217-6 (2011).
11      Ding, B., Turgeon, R. & Parthasarathy, M. V. Substructure of Freeze-Substituted Plasmodesmata. *Protoplasma* **169**, 28-41, doi:Doi 10.1007/Bf01343367 (1992).
12      Tilsner, J., Nicolas, W., Rosado, A. & Bayer, E. M. Staying Tight: Plasmodesmal Membrane Contact Sites and the Control of Cell-to-Cell Connectivity in Plants. *Annu Rev Plant Biol* **67**, 337-364, doi:10.1146/annurev-arplant-043015-111840 (2016).
13.     Nicolas, W. J. *et al.* Architecture and permeability of post-cytokinesis plasmodesmata lacking cytoplasmic sleeves. *Nat Plants* **3**, 17082, doi:10.1038/nplants.2017.82 (2017).
14      Roberts, A. G. & Oparka, K. J. Plasmodesmata and the control of symplastic transport. *Plant Cell Environ* **26**, 103-124, doi:DOI 10.1046/j.1365-3040.2003.00950.x (2003).

15      Burch-Smith, T. M., Stonebloom, S., Xu, M. & Zambryski, P. C. Plasmodesmata during development: re-examination of the importance of primary, secondary, and branched plasmodesmata structure versus function. *Protoplasma* **248**, 61-74, doi:10.1007/s00709-010-0252-3 (2011).

16      Schulz, A. Plasmodesmal Widening Accompanies the Short-Term Increase in Symplasmic Phloem Unloading in Pea Root-Tips under Osmotic-Stress. *Protoplasma* **188**, 22-37, doi:Doi 10.1007/Bf01276793 (1995).
17      Brunkard, J. O., Runkel, A. M. & Zambryski, P. C. The cytosol must flow: intercellular transport through plasmodesmata. *Curr Opin Cell Biol* **35**, 13-20, doi:10.1016/j.ceb.2015.03.003 (2015).
18      Zavaliev, R., Ueki, S., Epel, B. L. & Citovsky, V. Biology of callose (beta-1,3-glucan) turnover at plasmodesmata. *Protoplasma* **248**, 117-130, doi:10.1007/s00709-010-0247-0 (2011).
19      De Storme, N. & Geelen, D. Callose homeostasis at plasmodesmata: molecular regulators and developmental relevance. *Front Plant Sci* **5**, doi:10.3389/fpls.2014.00138 (2014).
20      Cui, W. & Lee, J. Y. Arabidopsis callose synthases CalS1/8 regulate plasmodesmal permeability during stress. *Nature Plants* **2**, doi:10.1038/Nplants.2016.34 (2016).

21      Fernandez-Calvino, L. *et al.* Arabidopsis Plasmodesmal Proteome. *Plos One* **6**, doi:10.1371/journal.pone.0018880 (2011).

22      Kraner, M. E., Muller, C. & Sonnewald, U. Comparative proteomic profiling of the choline transporter-like1 (CHER1) mutant provides insights into plasmodesmata composition of fully developed Arabidopsis thaliana leaves. *Plant J* **92**, 696-709, doi:10.1111/tpj.13702 (2017).
23      Diao, M. *et al.* Arabidopsis formin 2 regulates cell-to-cell trafficking by capping and stabilizing actin filaments at plasmodesmata. *Elife* **7**, doi:10.7554/eLife.36316 (2018).
24      Grison, M. S. *et al.* Specific membrane lipid composition is important for plasmodesmata function in Arabidopsis. *Plant Cell* **27**, 1228-1250, doi:10.1105/tpc.114.135731 (2015).

25	Zhang, Z. *et al.* Suppressing a Putative Sterol Carrier Gene Reduces Plasmodesmal Permeability and Activates Sucrose Transporter Genes during Cotton Fiber Elongation. *Plant Cell* **29**, 2027-2046, doi:10.1105/tpc.17.00358 (2017).

26	Ross-Elliott, T. J. *et al.* Phloem unloading in Arabidopsis roots is convective and regulated by the phloem-pole pericycle. *Elife* **6**, doi:10.7554/eLife.24125 (2017).

27	Vaten, A. *et al.* Callose biosynthesis regulates symplastic trafficking during root development. *Dev Cell* **21**, 1144-1155, doi:10.1016/j.devcel.2011.10.006 (2011).

28	Ullman M.D. & Radin N.S. The enzymatic formation of sphingomyelin from ceramide and lecithin in mouse liver. J Biol Chem 249: 1506–1512 (1974).

29	Elmendorf H.G. & Haldar K. Plasmodium falciparum exports the Golgi marker sphingomyelin synthase into a tubovesicular network in the cytoplasm of mature erythrocytes. J Cell Biol 124: 449–462(1994)

30	Tafesse, F. G., Ternes, P. & Holthuis, J. C. M. The multigenic sphingomyelin synthase family. *J Biol Chem* **281**, 29421-29425, doi:10.1074/jbc.R600021200 (2006).

31	Voelker, D. R. & Kennedy, E. P. Cellular and Enzymic-Synthesis of Sphingomyelin. *Biochemistry-Us* **21**, 2753-2759, doi:DOI 10.1021/bi00540a027 (1982).

32	Huitema, K., van den Dikkenberg, J., Brouwers, J. F. H. M. & Holthuis, J. C. M. Identification of a family of animal sphingomyelin synthases. *Embo J* **23**, 33-44, doi:10.1038/sj.emboj.7600034 (2004).

33	Yachi, R. *et al.* Subcellular localization of sphingomyelin revealed by two toxin-based probes in mammalian cells. *Genes Cells* **17**, 720-727, doi:10.1111/j.1365-2443.2012.01621.x (2012).

34	Wang, W. *et al.* An inositolphosphorylceramide synthase is involved in regulation of plant programmed cell death associated with defense in Arabidopsis. *Plant Cell* **20**, 3163-3179, doi:10.1105/tpc.108.060053 (2008).

35	Markham, J. E. & Jaworski, J. G. Rapid measurement of sphingolipids from Arabidopsis thaliana by reversed-phase high-performance liquid chromatography coupled to electrospray ionization tandem mass spectrometry. *Rapid Commun Mass Spectrom* **21**, 1304-1314, doi:10.1002/rcm.2962 (2007).

36	Markham, J. E., Li, J., Cahoon, E. B. & Jaworski, J. G. Separation and identification of major plant sphingolipid classes from leaves. *J Biol Chem* **281**, 22684-22694, doi:10.1074/jbc.M604050200 (2006).

37	Gronnier, J., Germain, V., Gouguet, P., Cacas, J. L. & Mongrand, S. GIPC: Glycosyl Inositol Phospho Ceramides, the major sphingolipids on earth. *Plant Signal Behav* **11**, e1152438, doi:10.1080/15592324.2016.1152438 (2016).

38	Axelrod, D., Koppel, D. E., Schlessinger, J., Elson, E. & Webb, W. W. Mobility measurement by analysis of fluorescence photobleaching recovery kinetics. *Biophys J* **16**, 1055-1069, doi:10.1016/S0006-3495(76)85755-4 (1976).

39	Merrill, A. H., Wang, E., Gilchrist, D. G. & Riley, R. T. Fumonisins and Other Inhibitors of De-Novo Sphingolipid Biosynthesis. *Adv Lipid Res* **26**, 215-234 (1993).

40	Markham, J. E. *et al.* Sphingolipids Containing Very-Long-Chain Fatty Acids Define a Secretory Pathway for Specific Polar Plasma Membrane Protein Targeting in Arabidopsis. *Plant Cell* **23**, 2362-2378, doi:10.1105/tpc.110.080473 (2011).

41	Benitez-Alfonso, Y. *et al.* Control of Arabidopsis meristem development by thioredoxin-dependent regulation of intercellular transport. *P Natl Acad Sci USA* **106**, 3615-3620, doi:10.1073/pnas.0808717106 (2009).

42	Li, W. M. *et al.* Depletion of ceramides with very long chain fatty acids causes defective skin permeability barrier function, and neonatal lethality in ELOVL4 deficient mice. *Int J Biol Sci* **3**, 120-128 (2007).

43  Schneiter, R. *et al.* A yeast acetyl coenzyme a carboxylase mutant links very-long-chain fatty acid synthesis to the structure and function of the nuclear membrane-pore complex. *Mol Cell Biol* **16**, 7161-7172 (1996).

44  Eisenkolb, M., Zenzmaier, C., Leitner, E. & Schneiter, R. A specific structural requirement for ergosterol in long-chain fatty acid synthesis mutants important for maintaining raft domains in yeast. *Mol Biol Cell* **13**, 4414-4428, doi: 10.1091/mbc.E02-02-0116 (2002).

45  Jean-luc C. *et al.* Revisiting Plant Plasma Membrane Lipids in Tobacco: A Focus on Sphingolipids. *Plant physiol* **170** 367–384 doi: 10.1104/pp.15.00564 (2016).

46. Wattelet-Boyer, V. *et al.* Enrichment of hydroxylated C24- and C26-acyl-chain sphingolipids mediates PIN2 apical sorting at trans-Golgi network subdomains. *Nat commun* **7** 12788 doi: 10.1038/ncomms12788 (2016).

47. Markham, J. E. *et al*. Sphingolipids containing very-long-chain fatty acids define a secretory pathway for specific polar plasma membrane protein targeting in Arabidopsis. *Plant cell* **23** 2362-2378 doi: 10.1105/tpc.110.080473 (2011).

48  Shang, B. S. *et al.* Very-long-chain fatty acids restrict regeneration capacity by confining pericycle competence for callus formation in Arabidopsis. *P Natl Acad Sci USA* **113**, 5101-5106, doi:10.1073/pnas.1522466113 (2016).

49  Clough, S. J. & Bent, A. F. Floral dip: a simplified method for Agrobacterium-mediated transformation of Arabidopsis thaliana. *Plant J* **16**, 735-743, doi: 10.1046/j.1365-313x.1998.00343.x (1998).

50  Wilke, S. A. *et al.* Deconstructing complexity: serial block-face electron microscopic analysis of the hippocampal mossy fiber synapse. *J Neurosci* **33** 507-522 doi:10.1523/JNEUROSCI.1600-12.2013 (2013).

51  Furuta, K. M. *et al.* Arabidopsis NAC45/86 direct sieve element morphogenesis culminating in enucleation. *Science* **345** 933-937 doi:10.1126/science.1253736 (2014).

52  Belevich, I., Joensuu, M., Kumar, D., Vihinen, H. & Jokitalo, E. Microscopy Image Browser: A Platform for Segmentation and Analysis of Multidimensional Datasets. *Plos Biol* **14**, doi:10.1371/journal.pbio.1002340 (2016).

53  Nicolas, W. J., Bayer, E. & Brocard, L. Electron Tomography to Study the Three-dimensional Structure of Plasmodesmata in Plant Tissues–from High Pressure Freezing Preparation to Ultrathin Section Collection. *Bio-protocol* 8(1): e2681. doi: 10.21769/BioProtoc.2681 (2018).

54  Kremer, J. R., Mastronarde, D. N. & McIntosh, J. R. Computer visualization of three-dimensional image data using IMOD. *J Struct Biol* **116**, 71-76, doi:10.1006/jsbi.1996.0013 (1996).

**Acknowledgements**

This work was supported by Finnish Centre of Excellence in Molecular Biology of Primary Producers (Academy of Finland CoE program 2014-2019, decision #271832),  the Gatsby Foundation (GAT3395/PR3), the National Science Foundation Biotechnology and Biological Sciences Research Council grant (BB/N013158/1), University of Helsinki (award 799992091) and the European Research Council Advanced Investigator Grant SYMDEV (No. 323052). This work was also funded as part of the DOE Joint BioEnergy Institute (http://www.jbei.org) supported by the U. S. Department of Energy, Office of Science,

Office of Biological and Environmental Research, through contract DE-AC02-05CH11231 between Lawrence Berkeley National Laboratory and the U. S. Department of Energy. We thank Raymond Wightman for assistance with confocal imaging, Matthieu Bourdon for help conducting immunolocalization and Katie Abley for seed germination assay. The SB-EM imaging was supported by Biocenter Finland (IB, EJ), and we would like to thank Mervi Lindman, Antti Salminen and Matias Veikkolainen for sample preparation for SB-EM. We thank Mary Roth and Ruth Welti for polar lipid analysis, Andrej Shevchenko and Kai Schuchmann for sphingomyelin analysis and Paul Dupree for data interpretation.

**Author information**

These authors contributed equally to this work: Dawei Yan, Shri Ram Yadav.

**Author notes**

William Nicolas

Present address: Division of Biology and Biological Engineering, California Institute of Technology, Pasadena, U. S. A

Affiliations

The Sainsbury Laboratory, University of Cambridge, Cambridge, United Kingdom

Dawei Yan, Andrea Paterlini, Ykä Helariutta

Helsinki Institute of Life Science / Institute of Biotechnology, University of Helsinki, 00014 Helsinki, Finland

Shri Ram Yadav, Eija Jokitalo, Ilya Belevich, Anne Vaten, Leila Karami, Sedeer el-Showk, Ykä Helariutta

Department of Biotechnology, Indian Institute of Technology, Roorkee, Uttarakhand, India

Shri Ram Yadav

Department of Plant and Soil sciences, Delaware Biotechnology Institute, University of Delaware, Newark, DE19711, U. S. A

Jung-Youn Lee

Biosciences Area, Lawrence Berkeley National Laboratory, CA, United States &

Joint Bioenergy Institute, Emeryville, CA, United States

Gosia M. Murawska, Jenny C. Mortimer

School of Biological Sciences, Washington State University, WA, United States

Michael Knoblauch

Department of Biochemistry, University of Nebraska-Lincoln, NE, United States

Jonathan E. Markham

Laboratory of Membrane Biogenesis, UMR5200 CNRS, University of Bordeaux, Bordeaux, France

William J. Nicolas, Magali S. Grison, Emmanuelle M. Bayer

Contributions

D.Y., A.P., S.R.Y., E.M.B. and Y.H. designed the experiments.  D.Y., A.P., S.R.Y. E.M.B and Y.H. wrote the manuscript with input from other authors. D.Y., S. R. Y.,

A.V. and J.Y.L generated Arabidopsis lines. S.R.Y., A.V.,L.K. and S.E.S carried out the suppressor screen and gene mapping. D.Y., S.R.Y. and A.P. performed growth phenotype analysis. D.Y., J.E.M. and M.S.G. carried out the sphingolipid profile. A.P. and W.J.N. carried out the electron tomography analysis with support from E.M.B.. M.K. provided support for FRAP experiment. A.P., E.J. and I.B. carried out the SB-EM analysis. G.M.M. and J.C.M. performed the GIPC glycosylation assay. The rest of experiments were performed by D.Y.

Competing interests

Corresponding author

Correspondence to Ykä Helariutta and Emmanuelle M. Bayer.

# Fig.1 Identification and characterization of *plm* mutants

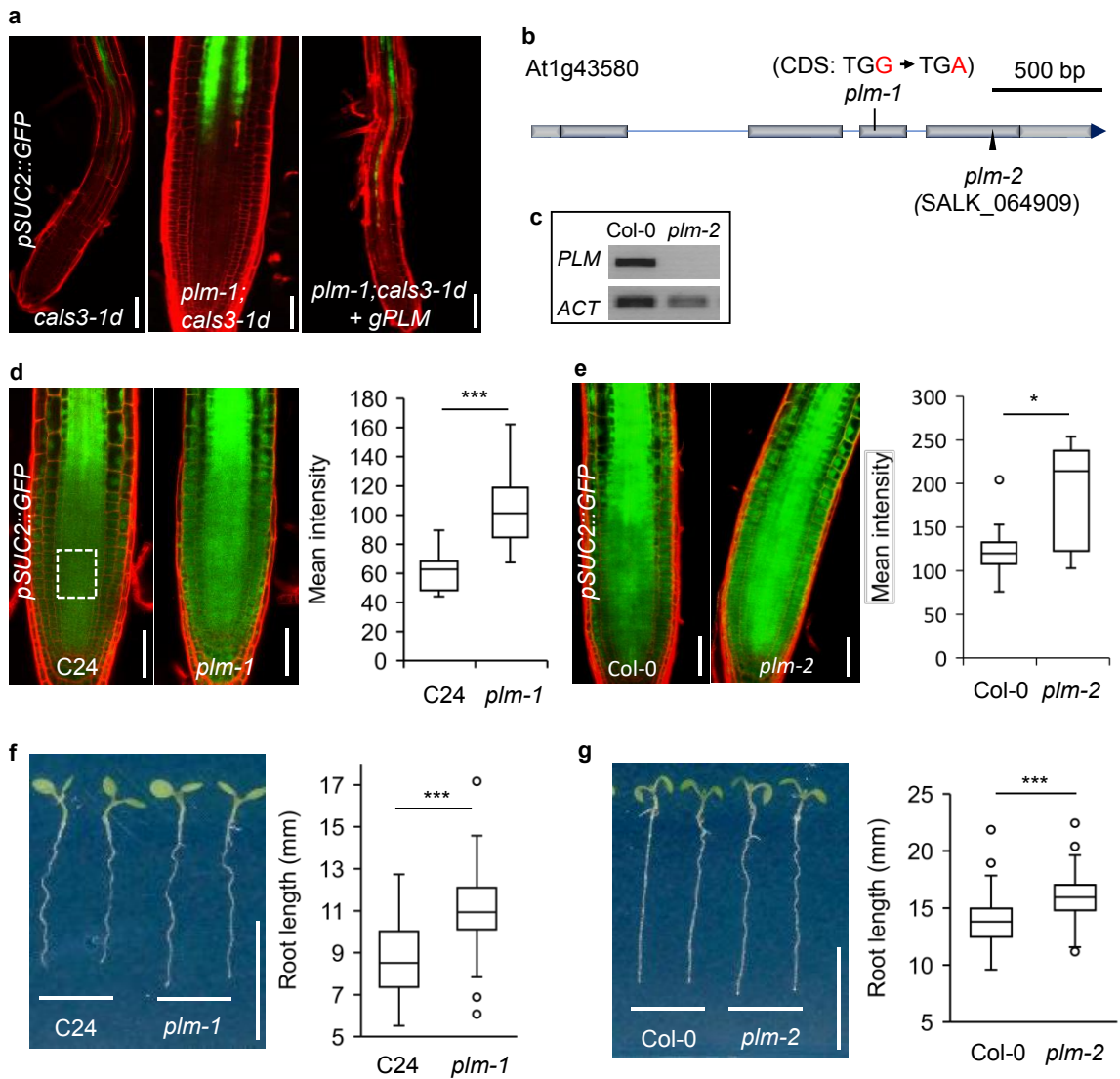

**a**. *pSUC2::GFP* in the roots of *cals3-1d*, *plm-1;cals3-1d* and *plm-1;cals3-1d* transformed with *PLM* genomic insert. **b**. Schematic view of the intron-exon structure of the *PLM* and sites of the mutations/T-DNA insertion. UTRs, white boxes. Exons, blue boxes. Introns, blue lines. **c**. RT-PCR showing undetectable transcript of *PLM* in *plm-2*. *ACTIN2* (*ACT*) was used as the internal control. **d**. Comparison of GFP intensity in the root meristems of wild-type C24 and *plm-1* mutant carrying *pSUC2::GFP*. An area of 50 x 40 µm$^2$ with 50 µm away from QC in the central root meristem (white dash box) was used for the quantification of fluorescence intensity, as well as all the following assays. n =16 (C24), n = 37 (*plm-1*). **e**. Comparison of GFP intensity in the root meristems of wild-type Col-0 and *plm-2* mutant carrying *pSUC2::GFP*. n =13 (Col-0), n = 16 (*plm-2*). **f**. Root length of 5-day-old wild-type C24 and *plm-1*. n=50 (C24), n = 46 (*plm-1*). **g**. Root length of 5-day-old wild-type Col-0 and *plm-2*. n=213 (Col-0), n = 205 (*plm-2*). Statistically significant differences were determined by Mann-Whitney-Wilcoxon test (*$p<0.05$, *** $p<0.001$). Scale bars, 50 µm (a,d,e), 1 cm (f), 10 cm (g).

**Fig.2 Molecular characterization of PLM**

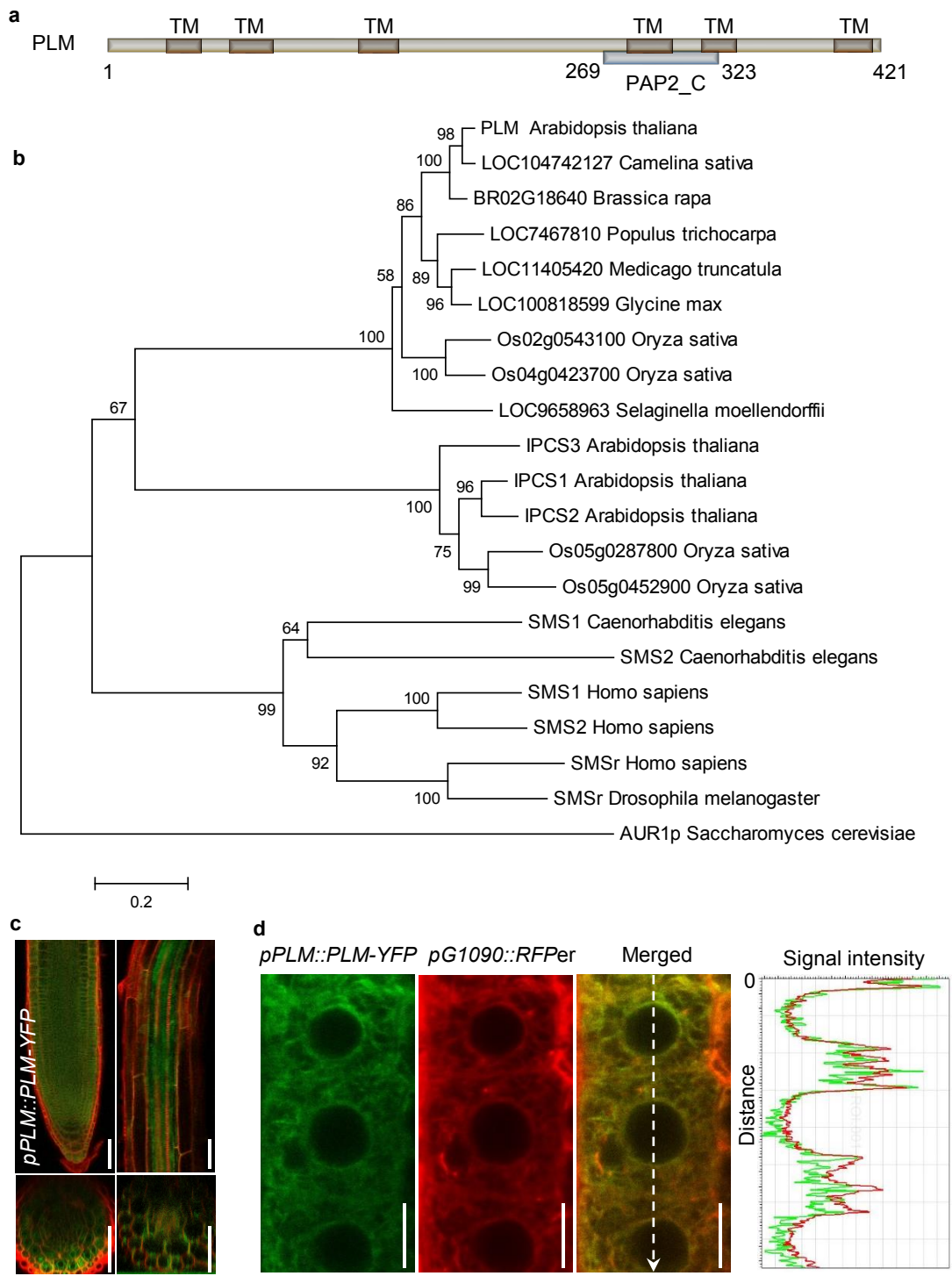

**a**. Schematic of PLM protein. TM, transmembrane domain. PAP2_C, type 2 phosphatidic acid phosphatase_C terminal. **b**. Phylogenetic tree of SMSs, IPCSs and PLM homologues from different organisms. Protein accession number and sequences are shown in Table 2. The tree was constructed by MEGA4.1 using neighbour-joining method. Bootstrap = 500. **c**. Expression pattern of *pPLM::PLM-YFP* in roots. **d**. Co-localization of *pPLM::PLM-YFP* with ER marker RFPer (with ER retention signal sequence HDEL). Signal intensities of GFP (green line) and RFP (red line) were obtained on the white. Scale bars, 50 μm (c), 10 μm (d).

# Fig.3 *PLM* affects sphingolipid biosynthesis.

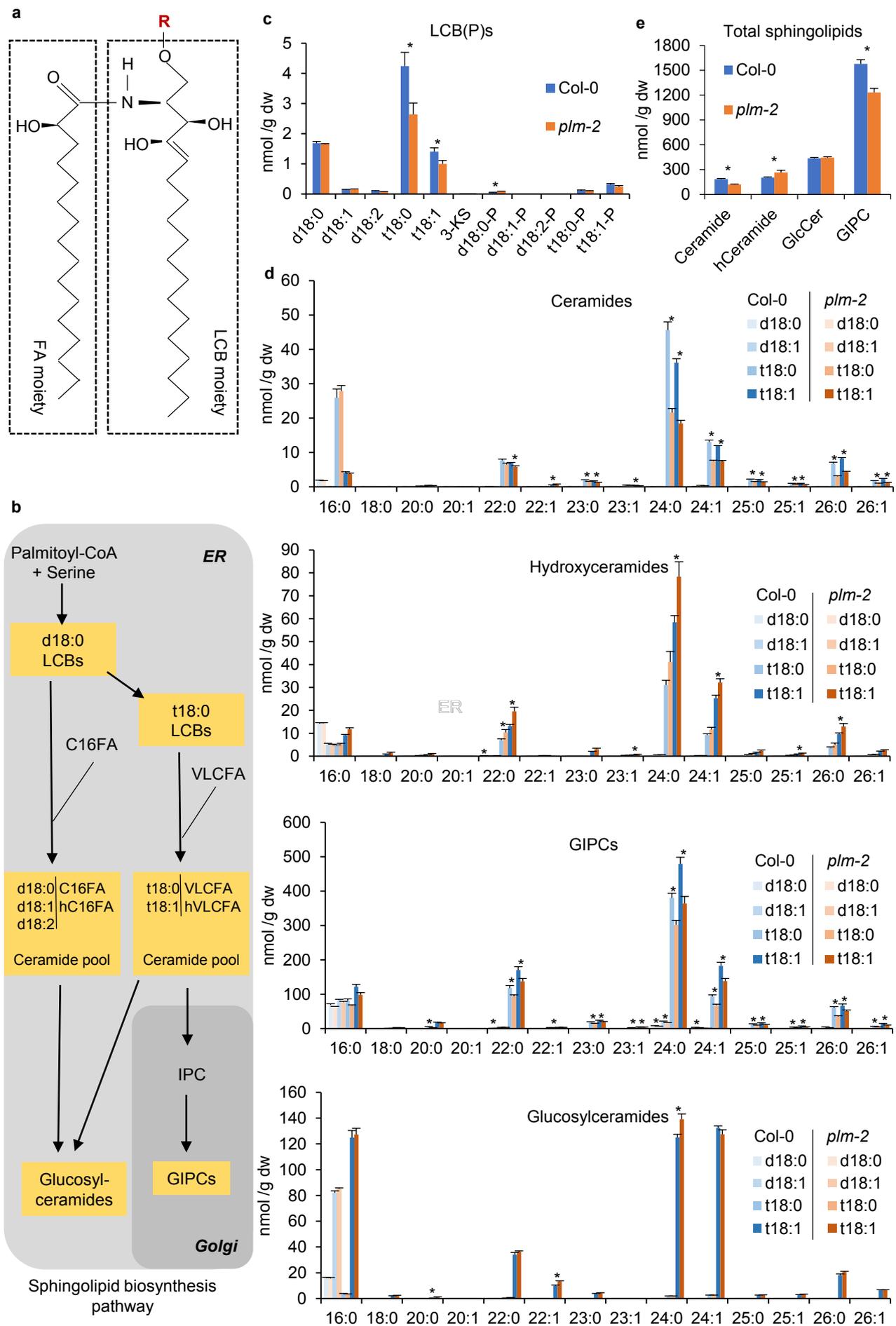

a. A diagram of sphingolipid structure. R indicates the substituent of sphingolipid. The FA chain has various carbon number while the LCB moiety are largely synthesized as saturated form (18:0) and desaturated form (18:1) with different OH (d/t) number. **b**. Simplified schematic representation of sphingolipid biosynthesis pathway in plants. **c**. Free LCB(P) levels in wild-type Col-0 and *plm-2*. **d**. Levels of complex sphingolipid species characterized by LCB(d18:0, d18:1, t18:0, t18:1) and FA(16:0-26:1) pairings in wild-type Col-0 and *plm-2*. **e**. Total levels of complex sphingolipid species in wild-type Col-0 and *plm-2*. Measurements are the average of five replicates. Statistically significant differences are indicated (two-tailed Student's *t*-test; *$p$<0.05).

# Fig. 4  GFP unloading into root tip is enhanced in *plm*

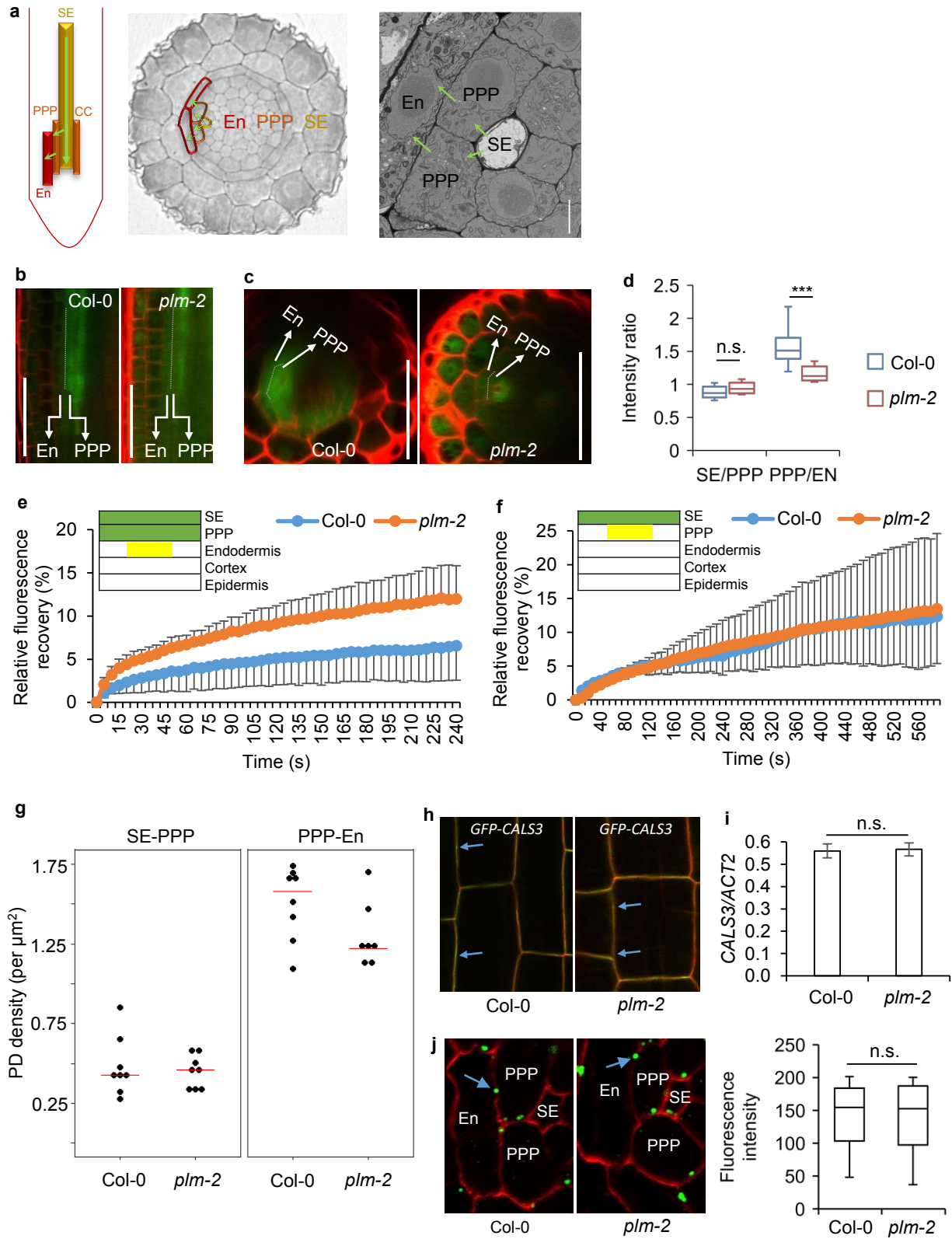

**a**. Schematic view of the unloading path of SE-PPP-endodermis in the root tip. **b**. GFP signals of PPP and endodermis at the unloading domains in the wild-type Col-0 and *plm-2* roots. **c**. Optical cross sections of (b). **d**. Fluorescence intensity ratios of SE/PPP and PPP/endodermis in the unloading domain (3 continuous cells since 200 µm away from QC post SE enucleation). n=13 (Col-0), n=11(*plm-2*). **e**. FRAP assay showing the recovery of GFP within the endodermis trafficking from PPP post bleaching of endodermis, cortex and epidermal layers in the unloading domain. The yellow box indicates the same region in endodermis described in (**d**) for GFP recovery assay. n=7 (Col-0), n=9 (*plm-2*). **f**. FRAP assay showing the recovery of GFP within the PPP trafficking from SE post bleaching of PPP, endodermis, cortex and epidermal layers in the unloading domain. The yellow box indicates the same region in PPP described in (**d**) for GFP recovery assay. n=4 (Col-0 and *plm-2* ). **g**. Plasmodesmata density assay by SB-EM at the SE-PPP and PPP-endodermis interface in the same region of unloading domain in Col-0 and *plm-2*. n=8 (Col-0), n=7 (*plm-2*). **h**. Localization of CALS3 (blue arrows) in Col-0 and *plm-2*. **i**. Relative transcript level of *CALS3* in Col-0 and *plm-2*. Data are based on three independent replicate experiments. The bars indicate average ± SD. **j**. Immunolocalization of callose around plasmodesmata (blue arrows) at the PPP-endodermis interface in the same region of unloading domain in Col-0 and *plm-2*. n=102 (Col-0) and 92 (*plm-2*) for quantification of fluorescence intensity. Significant differences were determined by Mann-Whitney-Wilcoxon test (d,j) and two-tailed student's *t*-test (i). *** *p*<0.001. n.s., no significance. Scale bars, 1 µm (**a**), 50 µm (**b**,**c**).

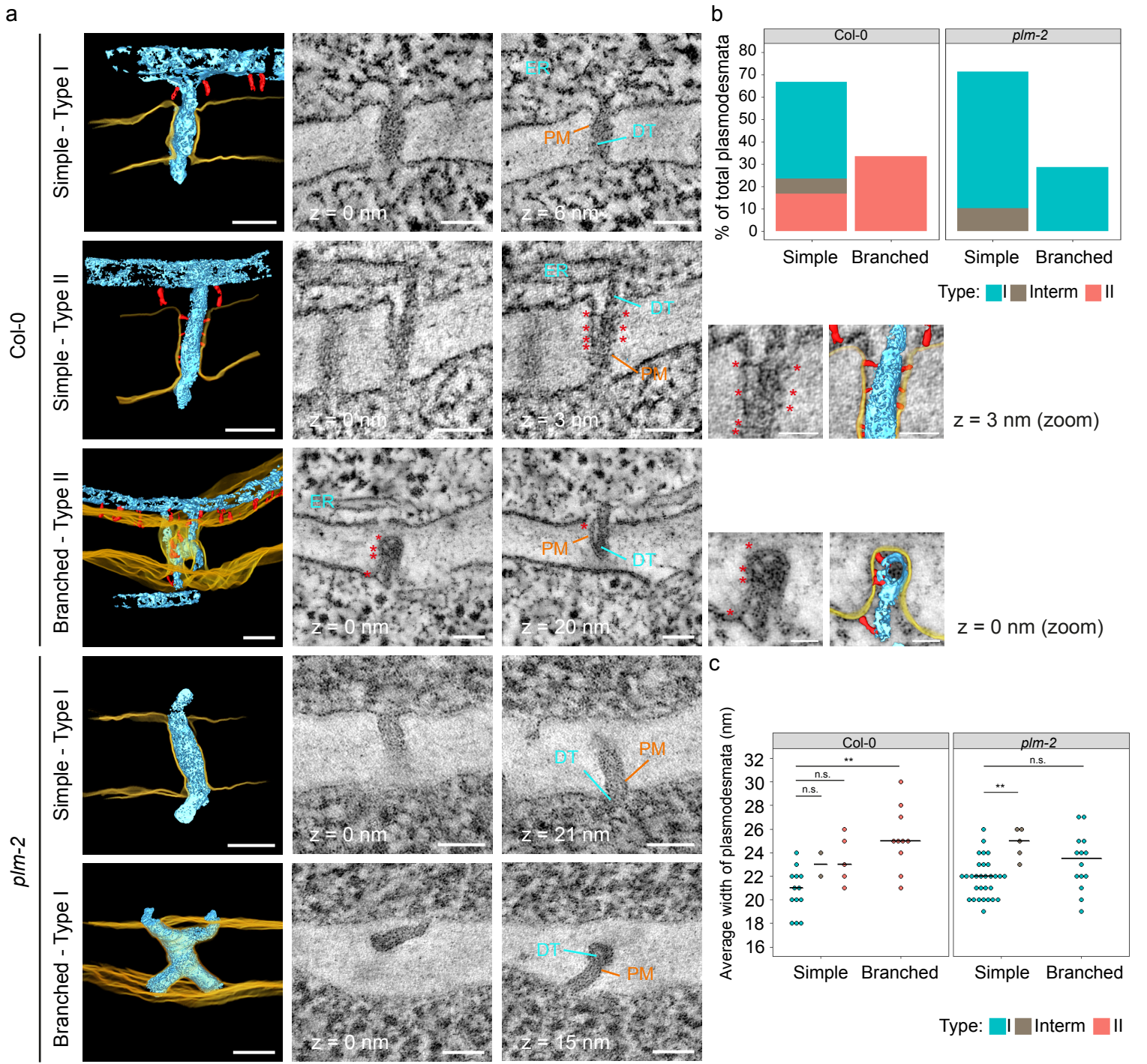

**Fig 5. *Plm-2* lacks type II plasmodesmata at the PPP-endodermal interface**

**a.** Plasmodesmata representing the various classes detected in Col-0 and *plm-2* roots with electron tomography at the PPP-endodermal interface, within the unloading domain. Panels with black background are 3D reconstructions based on 2D image stacks. Note that the 3D view of the branched plasmodesmatata in Col-0 is tilted relative to 2D images for ease of visualisation. Plasma membrane (PM) is rendered in yellow, desmotubule (DT) and ER in light blue, tethers within plasmodesmata and outside of them in red. Selected 2D views are displayed in the neighbouring panels with zooms for tethers within type II plasmodesmata (with a model overlay). Distance to the first 2D view is indicated under or next to each panel. Scale bars of 50nm (or 25 for zoomed panels) . Arrows, symbols and labels indicate various plasmodesmata components and follow the colour scheme of the 3D models (* = tether element within plasmodesmata).

**b.** Quantification of morphologies and types of plasmodesmata present at PPP-endodermal interface within the unloading domain in Col-0 and *plm-2* roots. n=30 (Col-0), n=49 (*plm-2*). The sections with plasmodesmata were acquired from two roots for each genotype.

**c.** Comparison of the widths of plasmodesmata described in panel b. Median value for each group is represented by black horizontal bar. Significant differences in reference to simple (type I) plasmodesmata in Col-0 or *plm-2* were determined by Dunn's test. The p-values were adjusted with Holm method for multiple comparisons. ** $p<0.01$, n.s. no significance.